# Über die Lichtgeschwindigkeit in bewegten Systemen[1]

## Norbert Feist

Leo-Fall-Str. 26
D 86368 Gersthofen

**Abstract.** Bradley beschrieb 1728 die jährliche Stern-Aberration. Respighi (1861) und Hoek (1868) zeigten als erste, daß eine solche bei terrestrischen Lichtquellen nicht existiert. Eine mögliche Deutung dieses Umstandes ist die, daß der Dualismus Welle/Teilchen schon in statu nascendi existiert: Bei einer bewegten Quelle könnte die Richtung der Normale der emittierten ebenen Welle wie bei einem ballistischen Modell trägheitsbedingt von der Geschwindigkeit der Quelle abhängen (siehe auch [1]). Die resultierende Kegelschnittgleichung für die Einweggeschwindigkeit $c_{(\phi)}=(c^2-v^2)/(c+v\cos\phi)$ und deren harmonischer Mittelwert für die Zweiweggeschwindigkeit $c_{(h)}=(c^2-v^2)/c$ für den Hin- und Rückweg im bewegten System werden vorgestellt. $\phi$ ist der Winkel zwischen der Emissions- und der Bewegungsrichtung der systemverbundenen Quelle. Die richtungsunabhängige Konstanz der Zweiweggeschwindigkeit wird an Hand vergleichbarer Experimente mit Licht- und Schallwellen diskutiert.

## Inhalt:

**0. Thesen – Folien 1 bis 4**

**1. Sechs Modelle zur Lichtgeschwindigkeit in bewegten Systemen**
I Klassisches Äthermodell (Anf. 20. Jh.)
II Spezielle Relativitätstheorie (Einstein 1905)
III Ballistisches Modell (Ritz 1908)
IV Pashsky – Relation (1918/1921)
V Marinov - Relation (1977)
VI Allgemeine Kegelschnitt - Relation (Feist 2000)

**2. Herleitung der allgemeinen Kegelschnitt - Relation**
2.1 Das Ergebnis des MM-Experiments: Die Zweiweglichtgeschwindigkeit
2.2 Verhältnisse auf dem transversalen Interferometerarm
2.2.1 Reflexion senkrecht zur Fahrtrichtung am bewegten 45-Grad-Spiegel
2.2.2 Emission einer bewegten Quelle senkrecht zur Fahrtrichtung
2.3 Emission einer bewegten Quelle unter beliebigem Winkel zur Fahrtrichtung

**3. Optische und akustische Experimente zur Kegelschnitt - Relation**
3.1 Experimente zur optischen Zweiweggeschwindigkeit
3.1 .1 Die Richtungsunabhängigkeit der optischen Zweiweggeschwindigkeit
3.1.2 Die Geschwindigkeitsabhängigkeit der optischen Zweiweggeschwindigkeit
3.2 Experimente zur akustischen Zweiweggeschwindigkeit in Luft
3.2.1 Die Richtungsunabhängigkeit der akustischen Zweiweggeschwindigkeit
3.2.2 Die Geschwindigkeitsabhängigkeit der akustischen Zweiweggeschwindigkeit

**4. Ausblick**

---

[1] With title *The propagation of light and sound in moving systems* and a new abstract in English submitted to the Los Alamos e-print archive arXiv, subject classification General Physics (April 2001).

# Über die Lichtgeschwindigkeit in bewegten Systemen *)

**These: Licht und Schall breiten sich in ihren Medien ( Äther / Luft ) wie folgt aus:**

- **Einweggeschwindigkeit** $c_\phi = \dfrac{c^2 - v^2}{c + v \cos \phi}$   ( = Kegelschnittrelation )

    v = Absolutgeschwindigkeit von System und Quelle

    $\phi$ = Winkel zwischen Emissions- und Bewegungsrichtung

- **Zweiweggeschwindigkeit** $c_h = \dfrac{c^2 - v^2}{c}$  (entspr. den Experimenten )

- **Eine Längenkontraktion in Bewegungsrichtung gibt es nicht.**

*) Ausführliche Manuskripte beim Autor; pdf–Datei bei   Norbert.Feist@ncsgmbh.de

# Optisches Michelson – Morley - Experiment

**Derzeitige Thesen**

- **Der „Senkrechte" Strahl findet den davoneilenden Spiegel durch Strahldivergenz oder leichte Dejustage von Quelle oder Strahlteiler.**

- **Das Nullergebnis ist für den Fall v = konst. durch die Lorentzkontraktion erklärbar.**

- **Es gibt keinen Äther.**

**Antithesen**

- **Quer zur Fahrtrichtung reflektierte oder emittierte ebene Wellen finden den Spiegel aus dem gleichen Grund, aus dem es bei terrestrischen Quellen keine Aberration gibt, nämlich durch trägheitsbedingte Vorwärtsstrahlung.**

- **Es gibt ein bevorzugtes Bezugssystem. Die ellipsoidale Lichtausbreitung ist abhängig von v und ϕ gemäß der Kegelschnittrelation. ==Die Zweiweggeschwindigkeit ist isotrop und erklärt das Nullergebnis==.**

# Kennedy – Thorndike – Experiment

**Derzeitige Thesen**

- **Bei energieneutraler Geschwindigkeitsvariation im Gravitationsfeld gilt die SRT.**
- **Das Nullergebnis bei ungleichen Armlängen und variablem v ist erklärbar durch Längenkontraktion und Zeitdilatation gemäß SRT.**
- **Die Lichtausbreitung ist unabhängig von der Geschwindigkeit der Quelle immer = c. Bei Uhrensynchronisation nach Einstein ist auch die Einweglichtgeschw. immer = c.**

**Antithesen**

- **Das Sonnensystem bewegt sich gravitativ gebunden auf einer Keplerellipse. Die Erde bewegt sich zykloidenförmig im halbjährlichen Wechsel auf unterschiedlichen Seiten dieser Ellipse in unterschiedlichen Abständen vom Zentrum unterschiedlich schnell.**
- **Deshalb ist nicht der transversale Dopplereffekt sondern dessen Quadrat in Form der gravitativen Rotverschiebung zu berücksichtigen. Diese ergibt mit der hier vorgestellten Zweiweggeschwindigkeit eine Invarianz der Wellenlänge und erklärt das Nullergebnis des K – T – Experiments.**

# Akustisches Michelson – Morley – Experiment

## Derzeitige Thesen

- **Es geht anders aus als das optische Experiment.**
- **Mit seiner Hilfe könnte man die Geschwindigkeit einer Schallquelle ermitteln.**

## Antithesen

- **Es führt zum gleichen Nullergebnis wie das optische Experiment, weil nach eigenen Versuchen die Zweiweggeschwindigkeit wie beim Licht isotrop ist. Auch deren Abhängigkeit von der Systemgeschwindigkeit v gemäß der Kegelschnittrelation konnte im Versuch quantitativ bestätigt werden.**
- **Da die Annahme einer Lorentzkontraktion hier zur Erklärung wirklich unsinnig wäre, sollte man auf diesen Begriff auch in der Optik verzichten.**
- **Es spricht nichts mehr dagegen, sondern wieder einiges dafür, daß elektromagnetische Wellen wie alle anderen Wellen ein Medium zur Ausbreitung benötigen.**

## 1. Sechs Modelle zur Lichtgeschwindigkeit in bewegten Systemen

Die folgenden Grafiken sind Hodographen der Geschwindigkeit des von der bewegten Quelle emittierten Lichtes relativ zu dieser mit v = 0,6 c bewegten Quelle. **c** ist die Lichtgeschwindigkeit im ruhenden Medium. E ist der Ort der Emission, Q kennzeichnet den derzeitigen Ort von Quelle und Beobachter. $c_{(\phi)}$ ist die Einweglichtgeschwindigkeit im bewegten System. Die einfach zu überblickenden parallel zu **v** liegenden Vektoren von $c_{(\phi)}$, nämlich $c_{(0°)}$ und $c_{(180°)}$, sind rechts herausgezeichnet. Die entsprechenden absoluten Lichtgeschwindigkeiten als Summe von **v** und $c_{(\phi)}$ sind links herausgezeichnet. Rechts neben der Vektordarstellung sind die relevanten Formeln zur Berechnung von $c_{(\phi)}$ und der daraus resultierenden Zweiweglichtgeschwindigkeit $c_{(h)}$ ( harmonisches Mittel aus Hin- und Rückgeschwindigkeit) angegeben.

I   Klassisches Äthermodell (Anf. 20. Jh.), Abb. 1a

Das klassische Äthermodell sah die Lichtausbreitung als Wellenerscheinung mit der isotropen konstanten Geschwindigkeit c in einem ruhenden Medium analog z. B. dem Schall. Es scheiterte an der bis heute gängigen - hier aber bestrittenen - Vorstellung, daß die Emission einer bewegten Quelle der einer ruhenden gleicht. Entsprechend beschreibt die auf der Galileitransformation beruhende Gl. (1) die Phasengeschwindigkeit einer Quelle, die bei jedem Emissionsakt kurz ruht und isotrop emittiert. Diese realitätsferne Gleichung führt fälschlich zu einer Richtungsabhängigkeit der Zweiweggeschwindigkeit in Gl. (2), die durch die Michelson - Morley - Experimente mit systemverbundenen Quellen nicht bestätigt werden konnte.

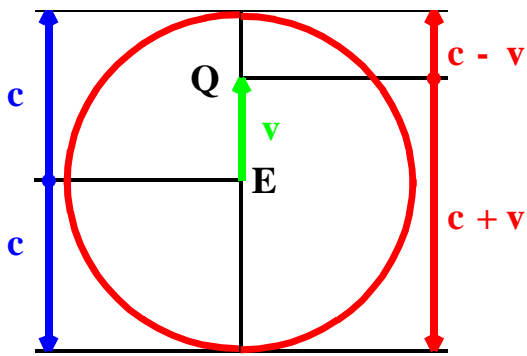

Abb. 1a Klassisches Äthermodell (19. Jh.)

$$c_{(\phi)} = \sqrt{c^2 - v^2 \sin^2 \phi} - v \cos \phi \quad (1)$$

$$c_{(h)} = \frac{c^2 - v^2}{\sqrt{c^2 - v^2 \sin^2 \phi}} \quad (2)$$

II   Spezielle Relativitätstheorie (Einstein 1905)

Das Modell der Speziellen Relativitätstheorie, das für den ruhenden wie für den bewegten Beobachter nur eine konstante Ein- und Zweiweglichtgeschwindigkeit c kennt, konnte sich etablieren.

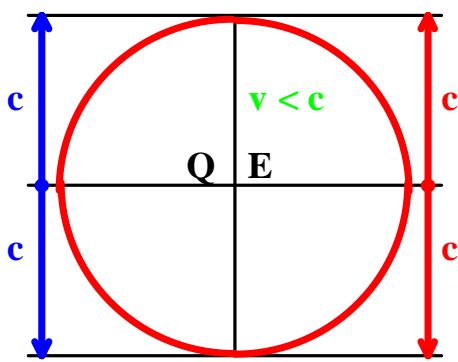

Abb. 1b Spezielle Relativitätstheorie (Einstein 1905)

$$c_{(\phi)} = \frac{c - v}{1 - \dfrac{v}{c}} = c \quad (3)$$

$$c_{(h)} = c \quad (4)$$

Die Notwendigkeit einer speziellen Gleichzeitigkeitsdefinition und der Annahme einer Längenkontraktion in Bewegungsrichtung sowie die Berücksichtigung der Frequenzabhängigkeit von der Geschwindigkeit („Zeitdilatation") erfordert für die Beschreibung der Vorgänge in einem anderen System eine aufwendige („Lorentz"-) Transformation für Weg und Zeit

$$x' = \frac{x - vt}{\sqrt{1 - \frac{v^2}{c^2}}} \qquad t' = \frac{t - (v/c^2)x}{\sqrt{1 - \frac{v^2}{c^2}}} \qquad (5)$$

die ihren Ausdruck auch in einem speziellen Additionstheorem der Geschwindigkeiten findet, s. Gl. (3). Zusammen mit dem Verzicht auf Äther, Absolutheit und Anschaulichkeit wird immer wieder die Vermutung genährt, daß die wahre Physik eine andere sein könnte. Es hat daher nie an Kritik gefehlt, allenfalls an umfassend konstruktiver. Eine solche bilden die nachfolgenden Modelle.

III  Ballistisches Modell (Ritz 1908)

Das ballistische Modell ging im Zuge der Photonenhypothese und des Dualismus Welle/Teilchen davon aus, daß sich die Photonen in einem Inertialsystem hinsichtlich Richtung und Betrag ihres Geschwindigkeitsvektors wie Korpuskeln verhalten. Die richtungsunabhängige Zweiweggeschwindigkeit $c_{(h)}=c$ hätte gut zum Ausgang des MM-Experiments gepaßt. Beobachtungen an Doppelsternsystemen zeigten jedoch, daß der Betrag des Lichtvektors nicht gleich **c**+**v** sein kann. Mit dem Scheitern des Modells gelangte auch der richtige Gedanke, daß die Richtung des Vektors von der Geschwindigkeit der Quelle abhängt, in den Mülleimer der Physikgeschichte.

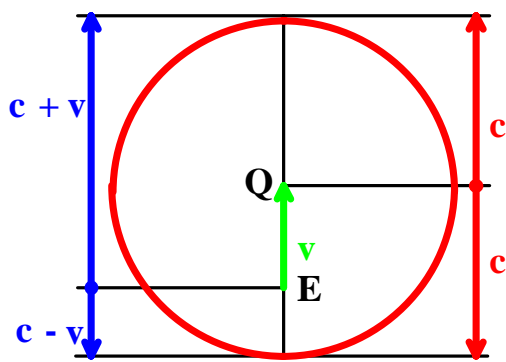

Abb.1c Ballistisches Modell (Ritz 1908)

$$\mathbf{c}_{(\phi)} = \mathbf{c} \qquad (6)$$

$$\mathbf{c}_{(h)} = \mathbf{c} \qquad (7)$$

IV  Pashsky – Relation (1918/1921) [11]

Pashsky sieht weder eine Längenkontraktion noch die Konstanz der Lichtgeschwindigkeit als notwendig oder erwiesen an. Ausgehend vom Ergebnis des MM – Experimentes fordert er lediglich die Isotropie der Zweiweggeschwindigkeit $c_h$ im bewegten System, deren Größe durchaus auch von der Geschwindigkeit v abhängen kann (wie in meinem Modell VI). Bezogen auf ein Ruhesystem sollten die Einweglichtgeschwindigkeiten einer bewegten Quelle parallel zur Bewegungsrichtung nicht c + v und c – v wie bei Ritz sondern c + $f_{1(v)}$ und c + $f_{2(v)}$ betragen mit der Maßgabe $f_{1(0)} = f_{2(0)} = 0$. Bezogen auf das bewegte System beträgt die Einweglichtgeschwindigkeit im Winkel $\phi$ zur Bewegungsrichtung

$$c_\phi = \frac{c_h}{1 + \sqrt{\frac{\Psi_{(v)}}{c_h + \Psi_{(v)}}} \cdot \cos\phi} \qquad (8)$$

Wenn $c_h$ und $\Psi$ beliebige Funktionen der Translationsgeschwindigkeit v sind mit der einzigen Bedingung, daß bei $v = 0$ $c_h = c$ und $\Psi = 0$ sind, so ergibt diese Relation „eine unbegrenzte Zahl von möglichen Hypothesen zur Erklärung des MM – Experimentes". In dieser unbegrenzten Zahl ist z. B. auch mein Modell VI enthalten, wenn man die Variablen wie folgt wählt:

$f_{1(v)} = f_{2(v)} = 0$ und $\Psi_{(v)} = \frac{v^2}{c^2}$.

Pashsky bzw. Rashevski, der nach dessen Tod die Ideen auf Wunsch der Witwe veröffentlichte, hoffte, daß die Experimentalphysik die präzise Form von $\Psi_{(v)}$ finden würde.

V  Marinov - Relation (1977) [2]

Marinov vertrat die korpuskulare Lichttheorie Newtons. Den beim MM – Experiment in Form einer isotropen Zweiweggeschwindigkeit aufgetretenen Widerspruch dazu beseitigte er wie Lorentz und Einstein mit der Annahme einer – allerdings nicht physischen – Längenkontraktion in Bewegungsrichtung und der Berücksichtigung der geschwindigkeitsabhängigen Frequenz von Quelle und Uhr für den Fall der Geschwindigkeitsänderung. Er setzte wie sie $c_h = c$, verzichtete aber auf Einsteins Vorschrift zur Uhrensynchronisation. Damit erweist sich die Einweglichtgeschwindigkeit im bewegten System als anisotrop entsprechend

$$c_\phi = \frac{c}{1 + \frac{v}{c} \cdot \cos\phi} \qquad (9)$$

und die Zeittransformation in der Lorentztransformation vereinfacht sich durch den Wegfall der räumlichen Koordinaten. Die so entstandene Marinov – Transformation zwischen absolutem und bewegtem System lautet:

$$x_2 = \frac{x - v_2 t}{\sqrt{1 - \frac{v_2^2}{c^2}}} \qquad t_2 = t\sqrt{1 - \frac{v_2^2}{c^2}} \qquad (10)$$

(Sie ist identisch mit einer 1961 von Tangherlini veröffentlichten Transformation. F. Selleri hat sie für den Fall zweier bewegter Systeme zur sogen. inertialen Transformation verallgemeinert [3]).

VI Allgemeine Kegelschnitt - Relation (Feist 2000)

Die allgemeine Kegelschnitt - Relation der Gl. (11) beschreibt aus Sicht eines mitbewegten Beobachters die Richtungs- und Geschwindigkeitsabhängigkeit der Phasengeschwindigkeit der Emission einer Quelle, die sich mit einem allseits offenen Inertialsystem geradlinig und konstant mit der traditionell als absolut bezeichneten Systemgeschwindigkeit **v** gegenüber einem ruhenden Trägermedium bewegt. Sie gilt für elektromagnetische Wellen im Vakuum ebenso wie für Schallwellen in Luft. Sie ergänzt die klassische Äthertheorie lediglich um die Annahme, daß die Richtung der Wellennormale emittierter ebener Wellen wie bei Korpuskeln trägheitsbedingt von der Geschwindigkeit der Quelle abhängt und erklärt damit das MM – Experiment ohne Annahme einer Längenkontraktion.

Bei der Transformation in ein System mit anderer absoluter Translationsgeschwindigkeit und/oder anderem Gravitationspotential sind bei elektromagnetischen Wellen die von Einstein beschriebenen realen Abhängigkeiten der Frequenz von Quelle und Uhr von der Geschwindigkeit bzw. dem Gravitationspotential als absolute nicht symmetrische Effekte zu berücksichtigen.

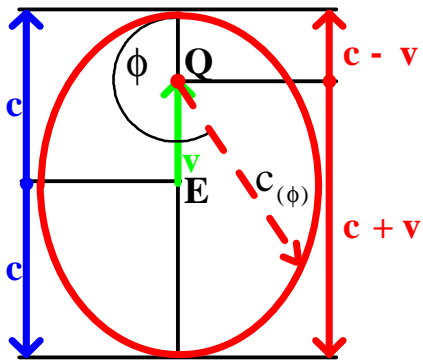

$$c_{(\phi)} = \frac{c^2 - v^2}{c + v \cos \phi} \quad (11)$$

$$c_{(h)} = \frac{c^2 - v^2}{c} \quad (12)$$

Eingezeichnet ist ein <u>beliebiger</u> Vektor $\mathbf{c}_{(\phi)}$.

Abb. 1d Allgemeine Kegelschnitt – Relation. Räumlich gesehen wäre der Hodograph der Geschwindigkeiten in bezug auf das bewegte System ein Rotationsellipsoid entlang der Hauptachse parallel zur Bewegungsrichtung mit der Quelle im Brennpunkt.

Man erkennt in der Grafik von Abb. 1d mit den Vektoren **c**, **v**, **c-v** und **c+v** die bei einer Emission parallel zu **v** für den gesunden Menschenverstand leicht zu überblickenden Verhältnisse des klassischen Äthermodells wieder. Ellipsoide und Ellipsen sind in der Natur nichts Fremdes: Man denke nur an die Bahnen von Elektronen oder Planeten. Mit v=0 ergibt sich die für das Ruhesystem angenommene isotrope kugelförmige Ausbreitung mit c.

Die Richtungsunabhängigkeit der Zweiweggeschwindigkeit von Gl. (12) entspricht dem Ergebnis des optischen MM-Experiments. Sie konnte - wie später (unter 3.2) ausgeführt - für Schallwellen ebenfalls experimentell bestätigt werden, ebenso wie die sich aus Gl. (12) ergebende Voraussage einer Abhängigkeit von der absoluten Systemgeschwindigkeit **v,** deren Bestätigung sich für den optischen Fall durch eine Neuinterpretation des Nullergebnisses von Kennedy – Thorndike – Experimenten im Gravitationsfeld ergibt (s. 3.1.2).

## 2. Herleitung der allgemeinen Kegelschnitt - Relation

Ausgangspunkt für die Herleitung der allgemeinen Kegelschnitt - Relation war eine Analyse und Neubewertung des Michelson - Morley - Experimentes (Abb. 2) mit Überlegungen zur Reflexion ebener Wellen an bewegten Spiegeln, zur Aberration und zur Trägheit des Lichtes. Beim MM-Experiment wurden im Michelsoninterferometer die Zweiweglichtgeschwindigkeiten $c_{(h)}$ in zwei zueinander senkrechten Richtungen miteinander verglichen.

2.1 Das Ergebnis des MM-Experiments: Die Isotropie der Zweiweglichtgeschwindigkeit

Das Ergebnis lautete: Die Zweiweglichtgeschwindigkeit ist entgegen der Erwartung auf beiden Armen gleich groß. Auf Grund der einfach zu überblickenden Verhältnisse parallel zu v

$$\frac{2}{c_{(h)}} = \frac{1}{c-v} + \frac{1}{c+v} \qquad (13)$$

$$c_{(h)} = \frac{c^2 - v^2}{c} \qquad (12)$$

ist anzunehmen, daß **diese** Zweiweggeschwindigkeit auch senkrecht zu v und letztlich für jede beliebige Richtung gilt.

2.2 Verhältnisse auf dem transversalen Interferometerarm

Es war zunächst zu klären, wie Gl. (12) in der zu v senkrechten Richtung erfüllt wird und zwar für die beiden physikalisch unterschiedlichen Fälle der <u>Reflexion</u> am halbdurchlässigen 45-Grad-Spiegel und der direkten <u>Emission</u> nach einer 90°-Drehung der Apparatur.

Abb. 2  Michelson – Interferometer, unterschiedliche Verhältnisse am jeweils orthogonalen Spiegel vor und nach der 90 – Grad – Drehung.

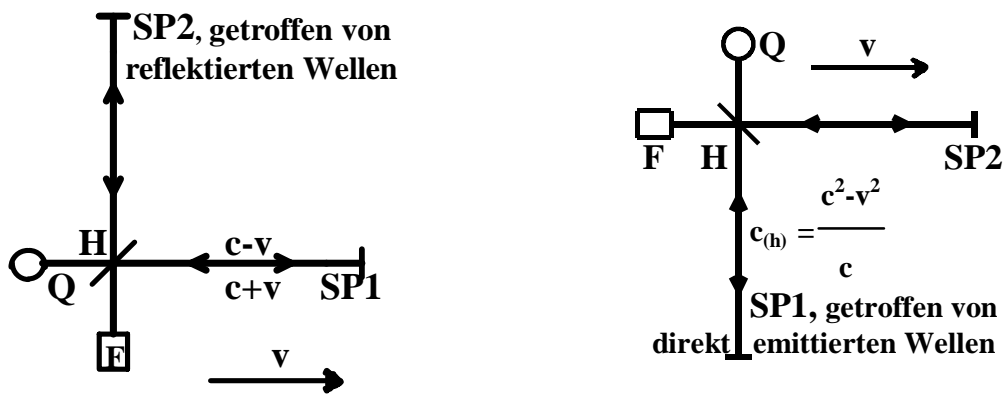

## 2.2.1 Reflexion senkrecht zur Fahrtrichtung am bewegten 45-Grad-Spiegel

Abb. 3a  Reflexion einer ebenen Welle (grün) am 45-Grad-Spiegel (blau) senkrecht zur Bewegungsrichtung.

Basislänge des 45-Grad-Spiegels = 4 cm, c = 3 cm/s, v = 1 cm/s, Frequenz $\nu$ = 1 / s, Wellenlänge $\lambda$ = 3 cm

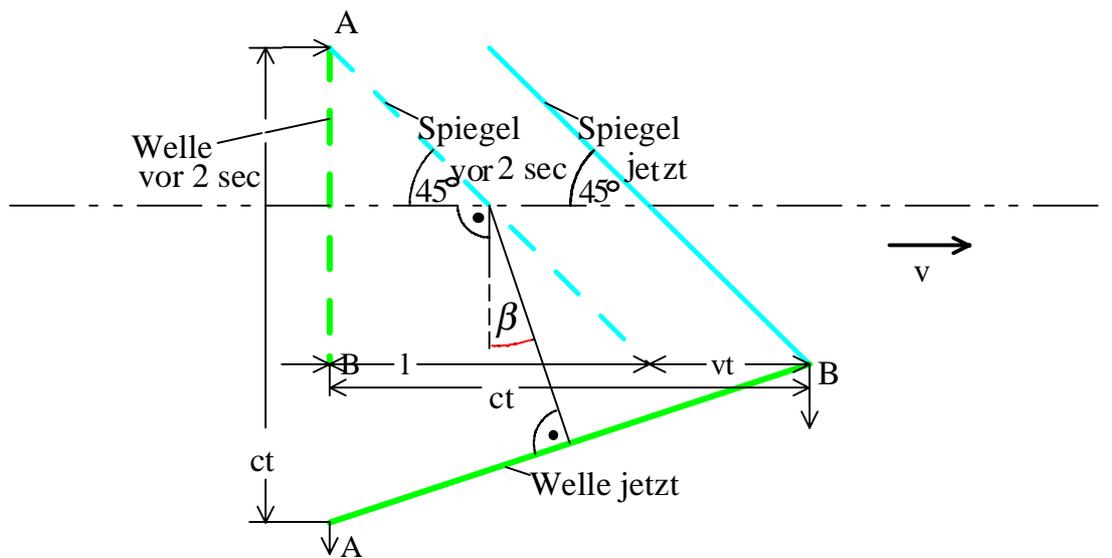

Abb. 3a zeigt aus Sicht eines nicht mitbewegten Beobachters strichliert das Auftreffen des Punktes A einer ebenen Welle und 2 Sekunden später durchgezogen das Ablösen des Punktes B der reflektierten Welle vom inzwischen um 2 cm davongeeilten Spiegel.

Die Dauer t der Reflexion berechnet sich nach der Gleichung

$$ct = l + vt \qquad (14)$$

Wesentlich ist, daß dadurch die senkrecht auf der Welle stehende Wellennormale um den Winkel

$$\beta = \arctan \frac{v}{c} \qquad (15)$$

in Fahrtrichtung geneigt wird.

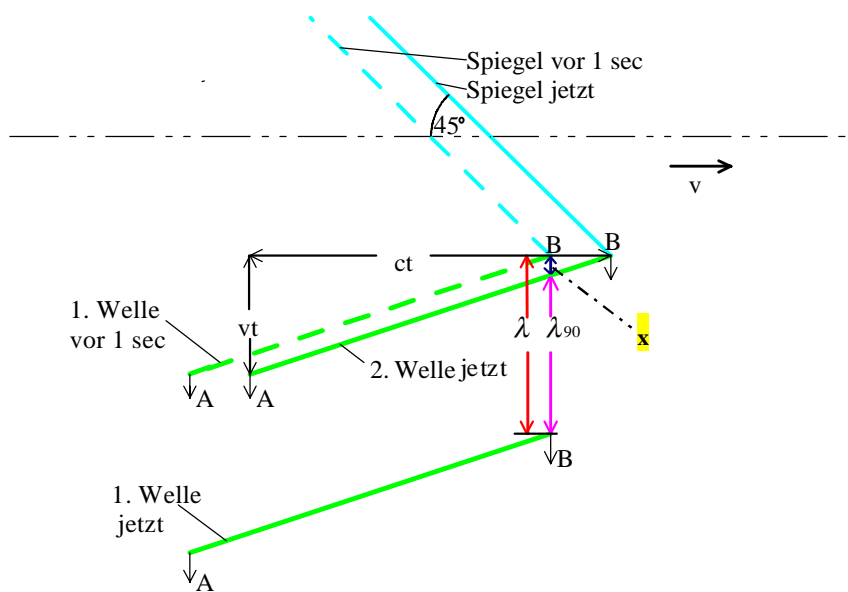

Abb. 3b  Reflexion am bewegten 45 – Grad – Spiegel entsprechend Abb. 3a

Die Abb. 3a wird in Abb. 3b ergänzt um die vorherige Welle, die ja eine Sekunde vorher den Spiegel verließ, der inzwischen die Strecke

$$\overline{BB} = \nu t v = \frac{c}{\lambda} t v \tag{16}$$

zurückgelegt hat. Die eingezeichnete Strecke vt ergibt sich aus Abb. 3a. Bedingt durch die Schrägstellung der Welle um den Winkel *β* und den räumlichen Versatz um die Strecke BB verringert sich in senkrechter Richtung der Abstand zweier Wellen um die Strecke x. Nach dem Strahlensatz wird die Wellenlänge in senkrechter Richtung zu

$$\lambda_{90} = \lambda(1 - \frac{v^2}{c^2}) \tag{17}$$

Multiplikation der Gleichung mit der im Inertialsystem der Quelle auch nach der Reflexion konstant bleibenden Frequenz ergibt die Einweglichtgeschwindigkeit

$$c_{(90)} = c(1 - \frac{v^2}{c^2}) = \frac{c^2 - v^2}{c}. \tag{18}$$

Nach Reflexion am Spiegel 2 des Interferometers folgt für den Rückweg die gleiche Einweggeschwindigkeit. Dann resultiert als harmonischer Mittelwert auf dem transversalen Arm des Interferometers ebenfalls die gleiche Zweiweggeschwindigkeit

$$\underline{\underline{c_{(h)} = \frac{c^2 - v^2}{c}}} \tag{12}$$

### 2.2.2 Emission einer bewegten Quelle senkrecht zur Fahrtrichtung

Abb. 4a Trägheitsgesetz bei „Emission" einer Korpuskel senkrecht zur Bewegungsrichtung

**Beispiel Ball:**

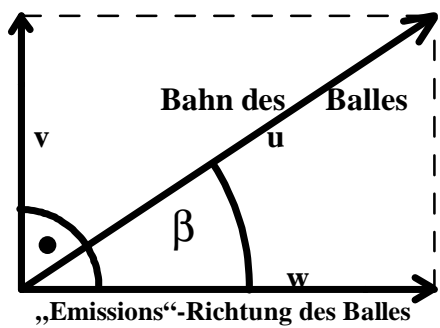

„Emissions"-Richtung des Balles

$$u = \sqrt{v^2 + w^2} \quad (19)$$

$$\tan\beta = \frac{v}{w} \quad (20)$$

Es ist anzunehmen, daß sich nicht nur bei der Reflexion, sondern auch bei der direkten Emission einer ebenen Welle senkrecht zur Fahrtrichtung deren Normale wie beim ballistischen Modell um den Winkel $\beta = \arctan\frac{v}{c}$ in Fahrtrichtung neigt. Wie in Abb. 4a und 4b gezeigt, soll also die Richtungskomponente des Lichtvektors dem Trägheitsgesetz gehorchen, ähnlich wie ein im beliebig schnell fahrenden Zug quer zur Fahrtrichtung geworfener Ball „automatisch" das gegenüberliegende Ziel trifft.

Wenn dem so ist, dürften mit dem bewegten System verbundene Quellen keine der Sternaberration analogen Effekte zeigen. Dies wird von der Literatur zwar nicht hervorgehoben, aber seit Respighis Untersuchungen von 1861 wiederholt bestätigt [4-7]. Auch die Vorwärtsrichtung bei Brems-, Synchrotron- oder Zerstrahlung sowie Darlegungen in [1] erhärten die Annahme $\beta = \arctan\frac{v}{c}$.

*[Bei der klassischen (m. E. nach unrichtigen) Herleitung des Michelson – Experiments oder bei Einsteins Lichtuhr geht man jedoch von der Aberration innerhalb eines bewegten Systems aus: Nicht der senkrecht emittierte Strahl findet aus Trägheitsgründen automatisch sein Ziel, sondern ein durch leichte Spiegeldejustierung abgelenkter oder ein glücklicherweise leicht schräg gerichteter Strahl aus dem „Strahlenbündel" trifft den davoneilenden orthogonalen Spiegel. Bei der Drehung um 180 Grad justiert entweder ein Kobold Spiegel oder Quelle in die jeweils andere Richtung oder die Divergenz der Quelle hält bequemerweise einen anders gerichteten Strahl bereit. Spätestens das optische Funktionieren der mit der Erde rotierenden km – langen Interferometer der Gravitationswellendetektoren mit den in engen evakuierten Rohren geführten und gefalteten Laserstrahlen bei einer Systemgeschwindigkeit von ca. 300 km/s sollte diesen Mechanismus ad absurdum führen. Siehe letzte Seite.]*

Abb. 4b Trägheitsgesetz bei Emission ebener Wellen senkrecht zur Bewegungsrichtung:

**Beispiel**

**ebene Welle:**

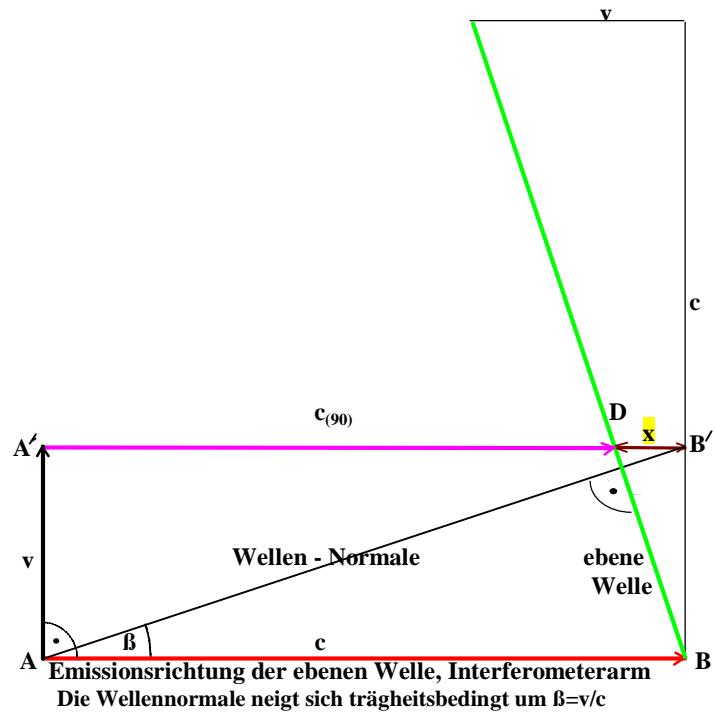

In der Abb. 4 b wurde bei A eine ebene Welle senkrecht zu v längs des Interferometerarmes in Richtung B emittiert (bei A' würde nach einer Sekunde die nächste Welle emittiert). Unter Annahme der trägheitsbedingten Neigung der Wellennormale um $\beta$ schneidet die ebene Welle den Interferometerarm nach einer Sekunde bei D, so daß für den mitbewegten Beobachter auf dem Interferometerarm die Einweggeschwindigkeit

$$c_{(90)} = c - x \qquad (21)$$

gilt. Nach dem Strahlensatz ist $\dfrac{x}{v} = \dfrac{v}{c}$ bzw. $x = \dfrac{v^2}{c}$, so daß wie im Reflexionsfall resultiert:

$$c_{(90)} = c_{(h)} = c - \frac{v^2}{c} = \frac{c^2 - v^2}{c} \qquad (22)$$

2.3 Emission einer bewegten Quelle unter beliebigem Winkel zur Fahrtrichtung

Abb. 5 Verhältnisse bei Emission unter beliebiger Richtung $\phi$ gegenüber **v**.
Bis auf die Größe der Winkel sind die Verhältnisse analog der Abb. 4.

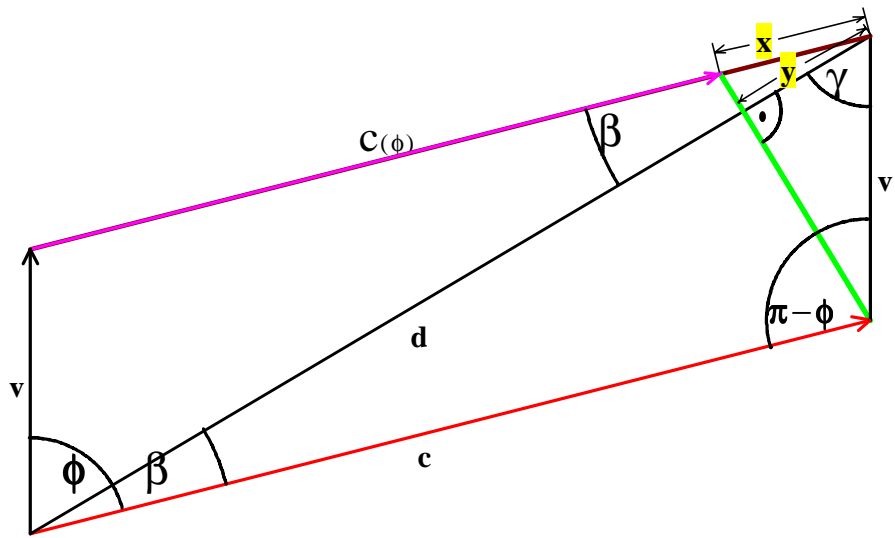

Dreimalige Anwendung des Kosinussatzes und die Substitutionen $\cos\gamma = \frac{y}{v}$ sowie $\cos\beta = \frac{y}{x}$ ergeben

$$y = \frac{v^2 + d^2 - c^2}{2d} \tag{23}$$

$$x = c\frac{v^2 + vc\cos\phi}{c^2 + vc\cos\phi} \tag{24}$$

und schließlich die allgemeine Kegelschnitt - Relation Gl. (9) der Einweggeschwindigkeit

$$c_\phi = \frac{c^2 - v^2}{c + v\cos\phi} \tag{11}$$

## 3. Optische und akustische Experimente zur allgemeinen Kegelschnitt – Relation

Im folgenden werden optische und akustische Experimente, die die Richtungsunabhängigkeit der Zweiweggeschwindigkeit und ihre Geschwindigkeitsabhängigkeit belegen, diskutiert bzw. vorgestellt.

## 3.1 Experimente zur optischen Zweiwegegeschwindigkeit

### 3.1.1 Die Richtungsunabhängigkeit der optischen Zweiweggeschwindigkeit

Die in Abb. 1d angegebene Gl. (12) enthält im Gegensatz zu Gl. (11) v ausschließlich in der zweiten Potenz. Das besagt, daß die Zweiweggeschwindigkeit eine richtungsunabhängige Systemkonstante ist.

Auf optischem Wege ist der Aspekt der Richtungsunabhängigkeit der Gl. (12) hinreichend oft durch das Michelson - Morley - Experiment in zahlreichen Ausführungsformen bestätigt worden. Infolge der richtungsunabhängigen Konstanz der Zweiweggeschwindigkeit bleibt - auch für den Fall ungleicher Armlängen - die Anzahl der Wellenlängen pro Arm bei beliebiger Drehung des Interferometers konstant. Daher kann es zu keiner Streifenverschiebung kommen.

### 3.1.2 Die Geschwindigkeitsabhängigkeit der optischen Zweiweggeschwindigkeit

Die Abhängigkeit der Zweiweglichtgeschwindigkeit von der Systemgeschwindigkeit v lässt sich durch passive Geschwindigkeitsvariation in einem Gravitationsfeld mit einem Kennedy – Thorndike – Experiment [12] indirekt bestätigen oder durch aktive Geschwindigkeitsvariation ähnlich dem Hafele - Keating – Experiment [13] direkt messen.

Nach Gl. (12) verringert sich die Zweiweggeschwindigkeit des Lichtes innerhalb des bewegten Systems, wenn dessen und der Quelle Geschwindigkeit $v_1$ auf $v_2$ erhöht wird, wie folgt:

$$c_{h1} = \frac{c^2 - v_1^2}{c} \quad \text{bzw.} \quad c_{h2} = \frac{c^2 - v_2^2}{c} \tag{12}$$

$$\Rightarrow c_{h2} = c_{h1} \frac{c^2 - v_2^2}{c^2 - v_1^2} \tag{25}$$

Die passive Bestätigung dieser Relation durch das Nullergebnis des K – T – Experiments erklärt sich durch die Anwendung der gravitativen Rotverschiebung. Der Schwerpunkt des Sonnensystems bewegt sich gravitativ gebunden auf einer Keplerellipse um einen anderen Schwerpunkt[*]. Sonne und Erde beschreiben dabei zykloidenförmige Bahnen analog den Bahnen von Erde und Mond (s. Experimental-Folie 2 im Anhang) um die Sonne. Auf der zentrumsfernen Seite der Bahn ist die Geschwindigkeit $v_1$ der Erde kleiner als $v_2$ ein halbes Jahr später auf der zentrumsnahen Seite.

Für die Frequenzabhängigkeit vom Gravitationspotential hatten Pound und Rebka [14] für den statischen Fall Einsteins Vorhersage der gravitativen Rotverschiebung

$$\nu_1 = \nu_0 \left(1 - \frac{MG}{Rc^2}\right) \tag{26}$$

experimentell bestätigt. Aus der Gleichheit von Gravitations- und Fliehkraft

$$\frac{mMG}{R^2} = \frac{mv^2}{R} \tag{27}$$

resultiert

$$\frac{MG}{R} = v^2 \tag{28}$$

---

[*] Dies gilt im Prinzip ebenso für eine noch übergeordnetere Bahnbewegung in der heutigen Dipolrichtung des Mikrowellenhintergrundes.

bzw. für den dynamischen Fall der Frequenzabhängigkeit vom Gravitationspotential in Gl.(26) eingesetzt

$$\nu_1 = \nu_0 \left(1 - \frac{v_1^2}{c^2}\right) \qquad \text{bzw.} \qquad \nu_2 = \nu_0 \left(1 - \frac{v_2^2}{c^2}\right) \qquad (29)$$

$$\Rightarrow \nu_2 = \nu_1 \frac{c^2 - v_2^2}{c^2 - v_1^2} \qquad (30)$$

Teilt man Gl. (25) durch Gl. (30), so zeigt das Ergebnis

$$\lambda_2 = \lambda_1, \qquad (31)$$

daß die Wellenlänge bewegter Quellen gegenüber derartigen Geschwindigkeitsänderungen invariant ist. Die interferometrisch durch K-T – Experimente sicher bestimmte Konstanz der Wellenlänge bei passiver gravitativer Änderung der Systemgeschwindigkeit v ist ein starker indirekter Beweis für die Geschwindigkeitsabhängigkeit der Zweiweggeschwindigkeit nach Gl. (12).

Deren Gültigkeit sollte aber bei praktisch konstantem Gravitationspotential auch durch aktive Geschwindigkeitsänderung mittels Energiezufuhr zum System direkt quantitativ messbar sein. Der Änderung der Zweiweglichtgeschwindigkeit

$$c_h = \frac{c^2 - v^2}{c} = c\left(1 - \frac{v^2}{c^2}\right) \qquad (12)$$

steht beim hier anzuwendenden transversalen Dopplereffekt lediglich eine Abhängigkeit der Uhrenfrequenz gemäß $\sqrt{1 - \frac{v^2}{c^2}}$ gegenüber. In einem nach Osten fliegenden Flugzeug sollte unter Berücksichtigung der rotationsbedingten Erdgeschwindigkeit am jeweilgen Breitengrad analog zum Hafele – Keating – Experiment eine vorausberechenbare kleinere Zweiweglichtgeschwindigkeit zu messen sein als bei einem Westflug.

3.2 Experimente zur akustischen Zweiweggeschwindigkeit in Luft

Wie einleitend erwähnt, gilt die allgemeine Kegelschnitt - Relation auch für Schallwellen in Luft. Zur Prüfung der Gl. (12) bezüglich Luftschall diente eine Meßstrecke bekannter Länge mit einem handelsüblichen Ultraschall - Entfernungsmesser und einem Aluminium - Reflektor an ihren Enden. Sie war zusammen mit einem schnellen Temperatursensor drehbar auf dem Dachgepäckträger eines PKW montiert, siehe Abb. 6.

Prinzip: Akustische Distanzmessung ( Puls/Echo - Laufzeit – Messung, Details in [15] )

Distanz: 1,35 m (ergibt bei 100 km/h + 8.7 mm, Standardabweichung im Stillstand 0,6 mm)

Links: Aluminiumblech 8 x 8,5 $cm^2$ als Reflektor

Rechts: Temperaturfühler mit Zeitkonstante 25 ms

Ganz rechts: 50 kHz-Folienwandler Format LR3S/242, Auflösung 0,1mm

Im PKW: Datenrekorder für Distanz, Geschwindigkeit und Temperatur

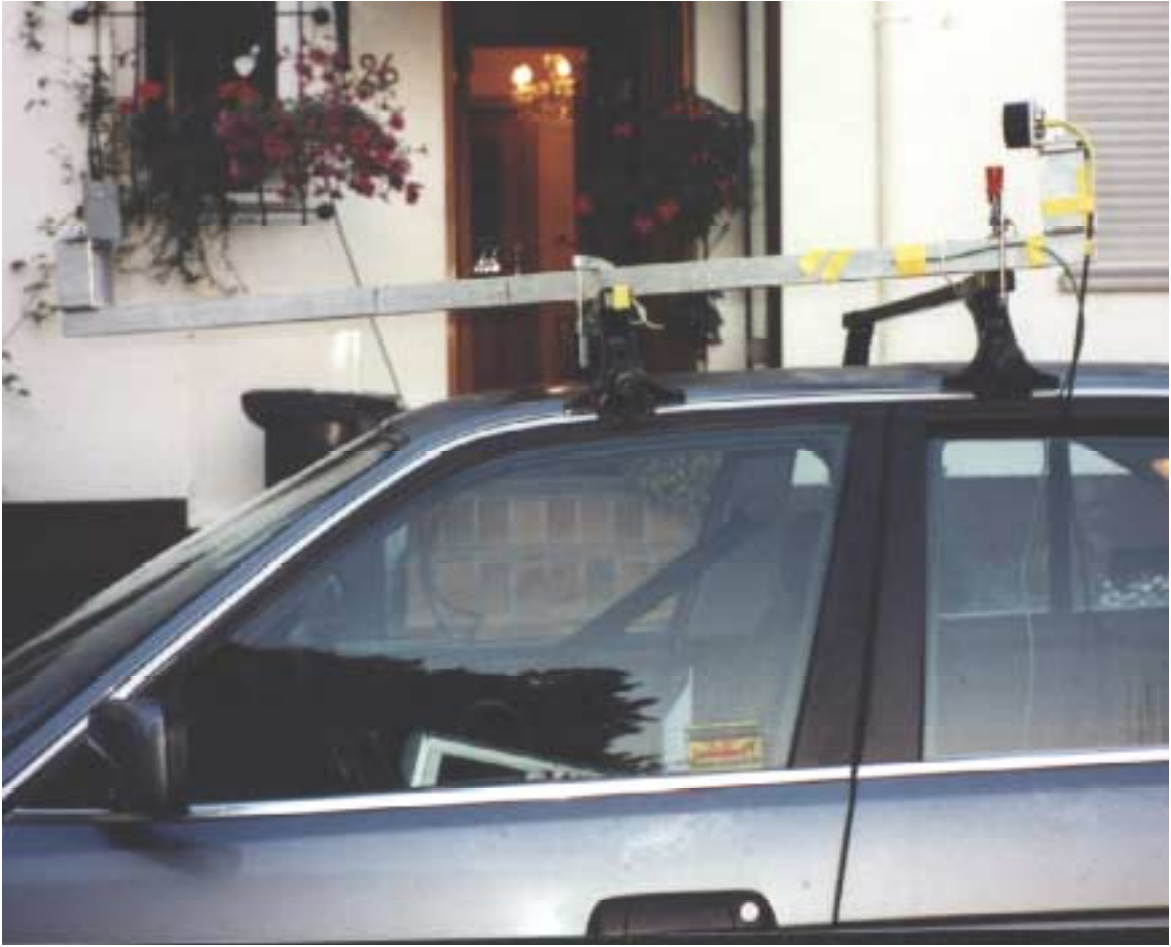

Abb. 6 Ultraschall – Laufzeit - Meßstrecke auf PKW

3.2.1 Die Richtungsunabhängigkeit der akustischen Zweiweggeschwindigkeit

Die Gl. (1) und (2) der klassischen Äthertheorie galten damals ebenfalls für Licht und Schall. Als die richtungsabhängige Gl. (2) durch den optischen MM-Versuch nicht bestätigt werden konnte, paßte hierzu gut Einsteins „Spezielle" Theorie für das Licht mit der Annahme c = const auch im bewegten System. Für den Schall sollten die Gleichungen (1) und (2) weiterhin gelten [8,9,10]. Die in Abb. 7 bis 12 dargestellten Ergebnisse Michelson-analoger Schallexperimente zeigen jedoch, daß diese genauso ausgehen wie beim Licht: Die Zweiweggeschwindigkeit im bewegten System ist richtungsunabhängig. In Abb. 7 - 12 werden z. B. für jeden Winkel bei v = 60 km/h als Zweiweggeschwindigkeit 99,75% von c erhalten. (Wie mir Dr. Karl Mocnik / Graz im Dezember 2000 mitteilte, hat er schon vor 10 Jahren festgestellt, daß Michelson - Experimente mit Schall genauso ausgehen wie die optischen.)

Abb. 7 - 12:   Meßergebnisse zur Richtungs- und Geschwindigkeitsabhängigkeit der akustischen Zweiweggeschwindigkeit:

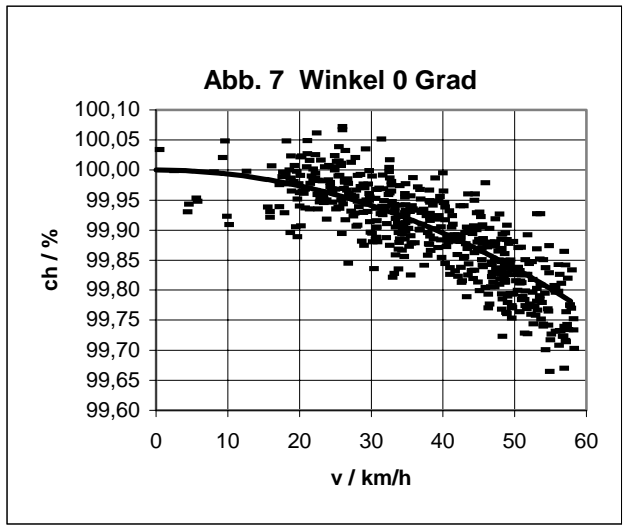

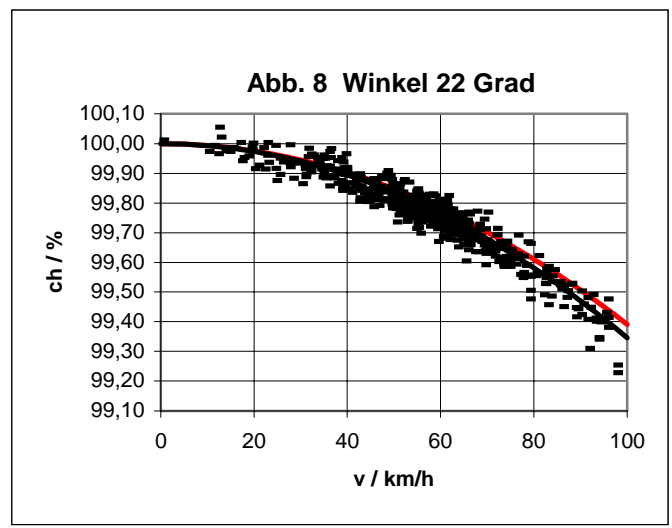

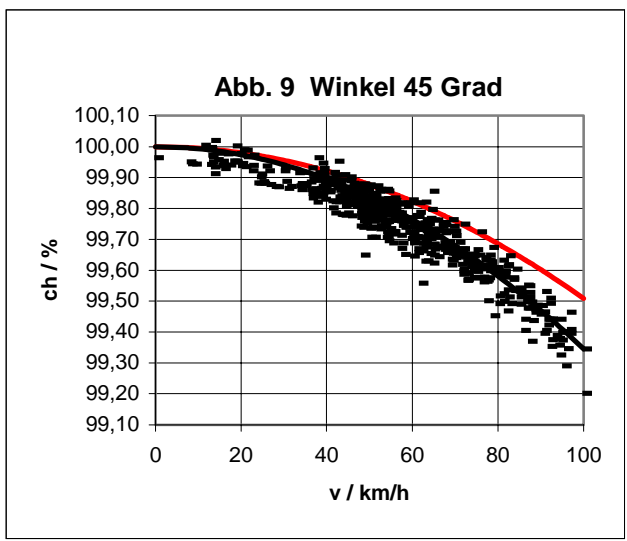

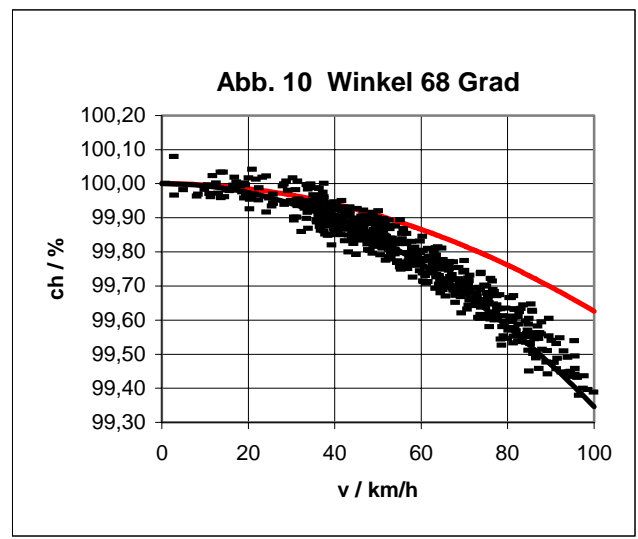

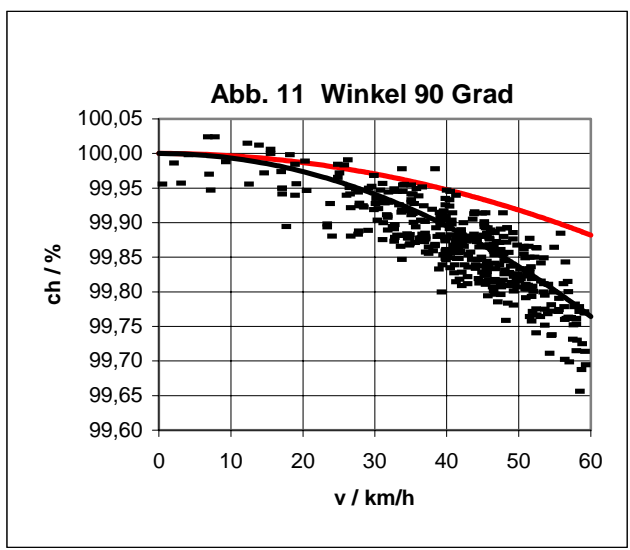

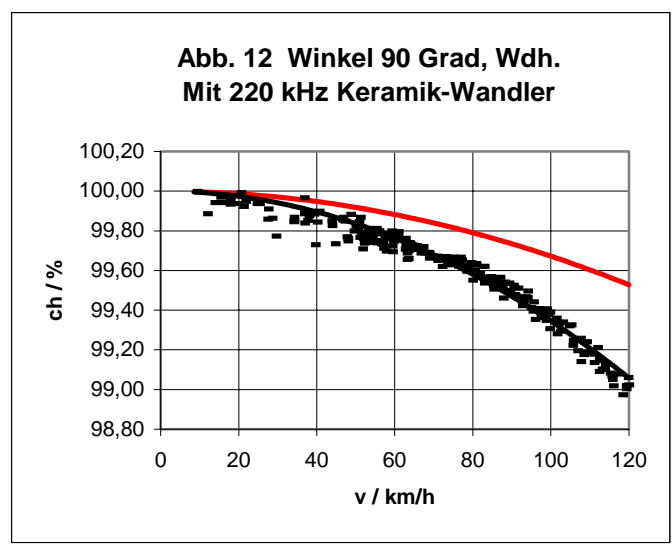

3.22 Die Geschwindigkeitsabhängigkeit der akustischen Zweiweggeschwindigkeit

Die allgemeine Kegelschnitt - Relation sagt mit Gl. (12) eine Geschwindigkeitsabhängigkeit der Zweiweggeschwindigkeit voraus. Wie in Abb. 7 - 12 ersichtlich, bestätigen die Meßwerte die für alle Winkel gleiche schwarze Sollwertkurve der Gl. (12) des Modells VI. Die rot eingezeichnete Sollkurve gemäß Gl. (2) der klassischen Äthertheorie (Modell I) weicht mit zunehmendem Winkel zunehmend ab. Bei Anwendung der Gleichungen der Modelle II, III und V (IV legt sich nicht fest) würde sich als Sollkurve eine waagerechte 100 % - Gerade ergeben, die noch stärker abweiche.

Wie ersichtlich, wird bisher nur die allgemeine Kegelschnitt - Relation durch die Ergebnisse der optischen und akustischen Experimente zur Zweiweggeschwindigkeit bestätigt.

# 4 Ausblick

Wenn - wie angenommen - die allgemeine Kegelschnitt - Relation für elektromagnetische Wellen zutrifft, kennen wir bisher auf Grund der angewandten Meßtechnik nur die hochgenau bestimmte Zweiweggeschwindigkeit des Lichtes

$$c_{(h)} = \frac{c^2 - v^2}{c} = 299792458{,}8 \pm 0{,}2 \, m/s \quad (32)$$

und nicht die Lichtgeschwindigkeit c. Sie läge mit

$$c = \frac{c_{(h)}}{2} \pm \sqrt{\left(\frac{c_{(h)}}{2}\right)^2 + v^2} \quad (33)$$

um ca. 300 m/s höher als dieser Wert, wenn die Annahme einer Absolutgeschwindigkeit v der Lokalen Gruppe in der Größenordnung von 300 km/s in Richtung des Sternbildes Löwe stimmt, wie sie die 180-Grad-Asymmetrie der Hintergrundstrahlung nahelegt. Die Einweglichtgeschwindigkeiten in unserem System wiesen dann logischerweise parallel zur Bewegungsrichtung einen maximalen Unterschied von ca. 600 000 m/s auf. Ich bin mir sicher, daß uns die genauen Werte nicht ewig verschlossen bleiben werden.

Abschließend noch ein Wort zur Absolutgeschwindigkeit v, die sich traditionell definitionsgemäß auf ein ruhendes Trägermedium, beispielsweise einen Äther, bezieht: Ergäbe eine örtliche räumliche Untersuchung der Geschwindigkeitsverteilung der Ätherteilchen, daß diese z. B. nach links immer um 20 km/s schneller fliegen würden als nach rechts, so bewegte sich das System mit der Absolutgeschwindigkeit v = 10 km/s nach rechts.

Ein anderes Inertialsystem mit der gleichen Zweiweggeschwindigkeit und damit auch der betragsmäßig gleichen Absolutgeschwindigkeit v wie das unsere bewegt sich also nicht unbedingt parallel mit uns von einem gemeinsamen räumlichen Zentrum weg. Es kann uns genausogut entgegen oder aus ganz anderer Richtung kommen. Deshalb zeigen z. B. auch zwei sich in Beschleunigern oder Speicherringen gegenläufig bewegende Teilchen den gleichen absoluten Effekt der Abhängigkeit ihrer Masse und Frequenz von ihrer betragsmäßig gleichen absoluten Geschwindigkeit v [15].

# Literatur

**Register der projizierten Overhead-Folien**

| | |
|---|---|
| Thesen-Folie 1 | Licht- und Schallausbreitung |
| Thesen-Folie 2 | MM - Experiment |
| Thesen-Folie 3 | KT - Experiment |
| Thesen-Folie 4 | MM und Schall |
| Modell-Folie 1 | Klassisches Äthermodell |
| Modell-Folie 2 | Spezielle Relativitätstheorie |
| Modell-Folie 3 | Ballistisches Modell |
| Modell-Folie 4 | Pashsky- und Marinov – Relationen |
| Modell-Folie 5 | Allgemeine Kegelschnitt – Relation |
| Ableitungs-Folie 1 | MM-Experiment: Reflexion und Emission |
| Ableitungs-Folie 2 | Reflexion am bewegten Spiegel 1 |
| Ableitungs-Folie 3 | Reflexion am bewegten Spiegel 2 |
| Ableitungs-Folie 4 | Ball senkrecht emittiert |
| Ableitungs-Folie 5 | Welle senkrecht emittiert |
| Ableitungs-Folie 6 | Emission unter beliebigem Winkel |
| Experimental-Folie 1 | Details zum KT – Experiment |
| Experimental-Folie 2 | Erde- Mond - Bahn |
| Experimental-Folie 3 | PKW mit Meßaufbau |
| Experimental-Folie 4 | Daten zur Meßstrecke |
| Experimental-Folie 5 | Abb. 7 - 12 |
| Experimental-Folie 6 | Abb. 12 groß |
| Experimental-Folie 7 | Original – Versuchsausdruck |

<u>Anhang</u>: Experimental-Folien 2 und 7, die nicht aus dem vorliegenden Text oder den Abbildungen extrahiert wurden sowie die Folie „Michelson – Interferometer als Gravitationswellendetektor". <u>Anmerkung zur Seitennummerierung</u>: Das Papier beginnt mit Seiten-Nr. 2.

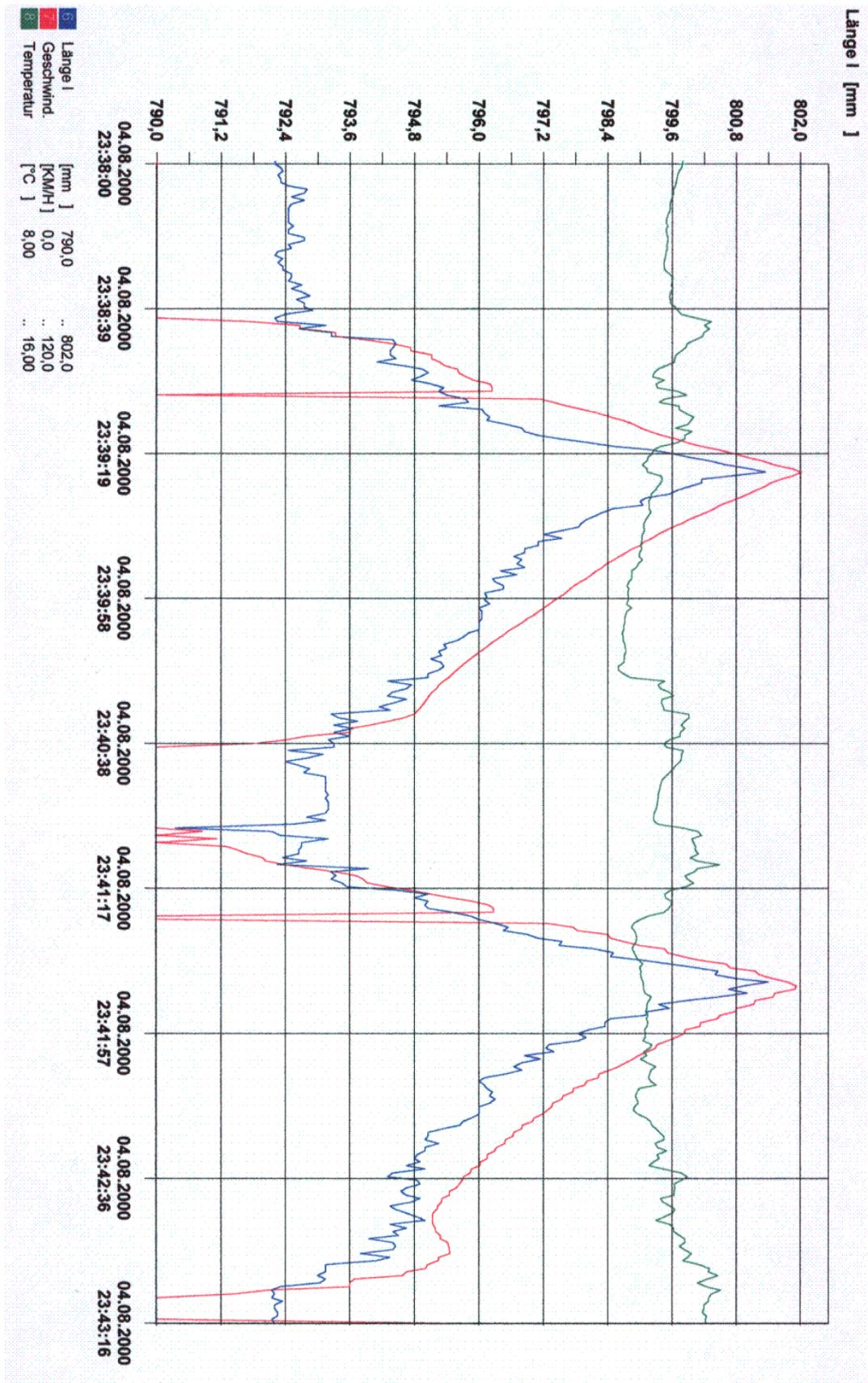

**Versuchsdaten-Ausdruck    Experimental-Folie 7**

# Erde - Mond – Bahn

Experimental-Folie 2

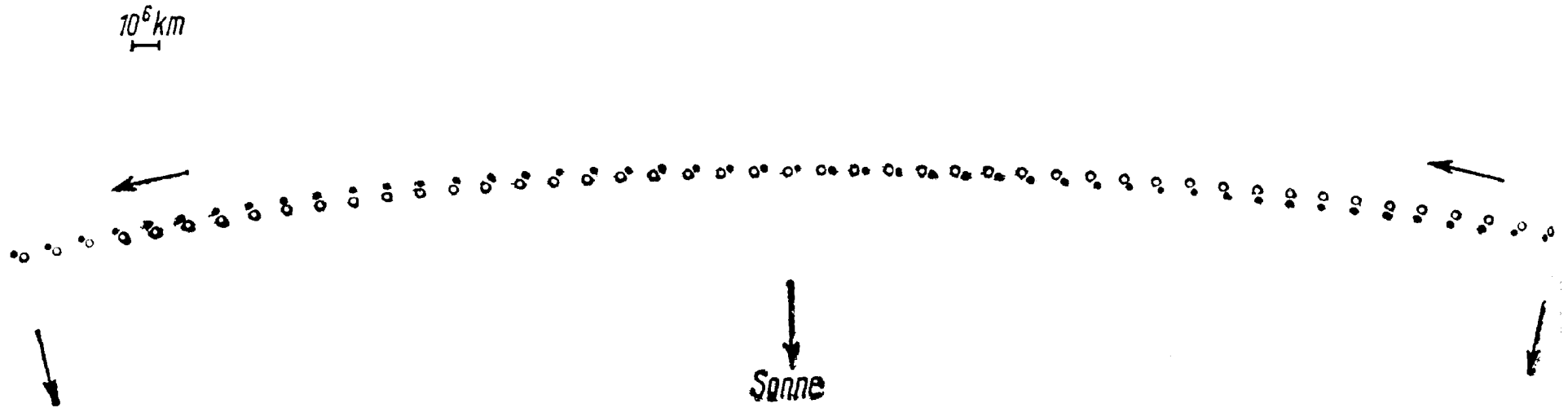

Ausschnitt der Bahn von Erde und Mond um die Sonne.
Positionen von Erde und Mond im Abstand von 0,5° heliozentrischer Länge (d. h. aller 12 h 10 min) während 23 Tagen

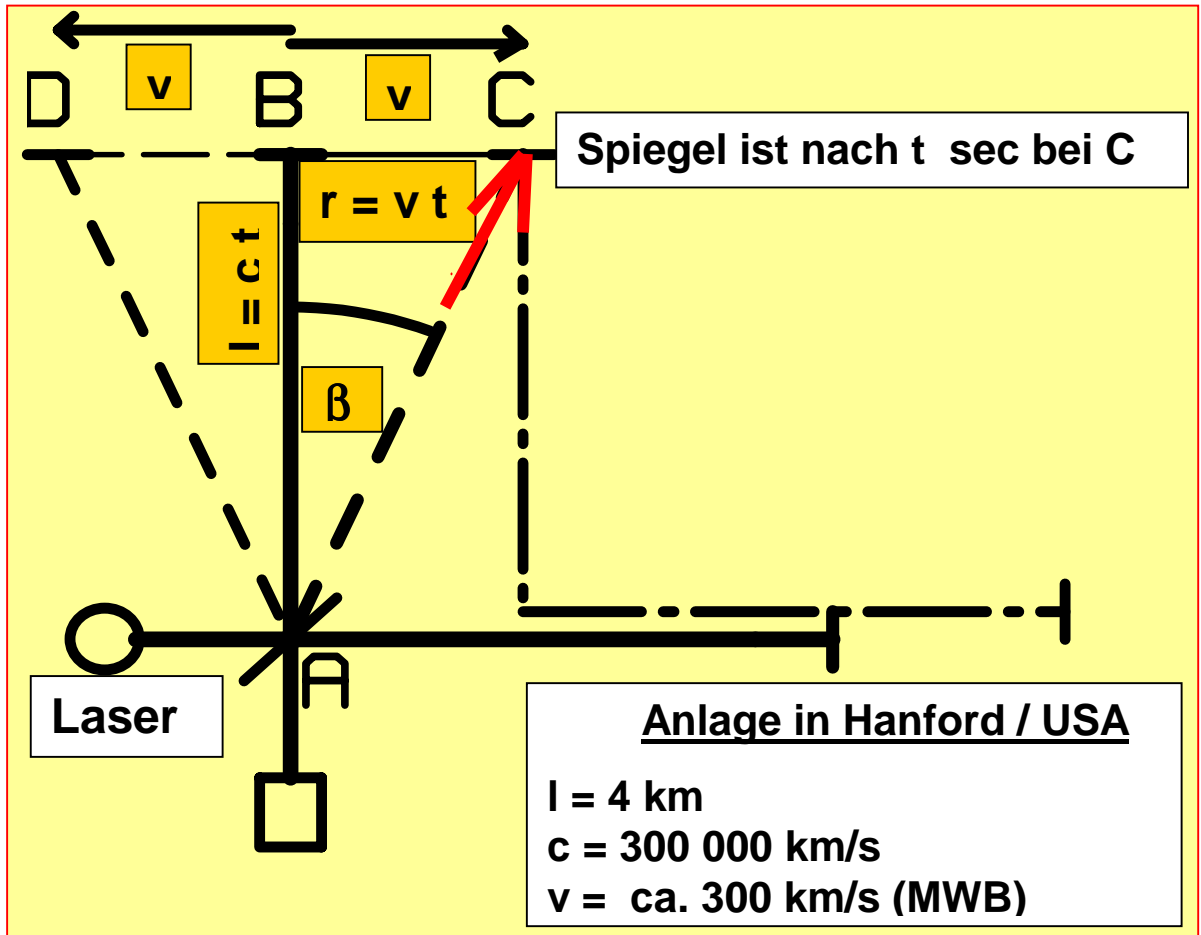

**Michelson - Interferometer als Gravitationswellendetektor: Wie trifft das Licht t sec nach der senkrechten Emission den Spiegel bei C ?   (Es ist   r : l = v : c     r = 4 m)**

**1) Durch Strahldivergenz**: Der Divergenzradius des Lasers müsste nach 4 km schon 4 m betragen. Der Laser der Monddistanzmessung erzeugt auf dem ca. 400 000 km entfernten Mond einen 20 km$^2$–Fleck. Er hat in 4 km Entfernung einen Radius von nur  r = 2,5 cm!

**2) Durch Spiegeldejustage bei A:** Ein Kobold müsste den halbdurchlässigen Spiegel **ständig nachjustieren**, so daß er nach 12 Std. oder halber Erdumdrehung (v entgegengesetzt) nach D weist.

**3) Korrekt: Durch trägheitsbedingte Vorwärtsrichtung des Lichtvektors. Deshalb gibt es auch in anderen Fällen keine Aberration bei terrestrischen Lichtquellen. So ist es auch beim MM-Experiment.**

25